\documentstyle[aps,epsf,aps10,amssymb]{revtex}
\catcode`\@=11
\def\lesssim{\mathrel{\mathpalette\vereq<}}
\def\vereq#1#2{\lower3pt\vbox{\baselineskip1.5pt \lineskip1.5pt
\ialign{$\m@th#1\hfill##\hfil$\crcr#2\crcr\sim\crcr}}}

\def\Let@{\relax\iffalse{\fi\let\\=\cr\iffalse}\fi}
\def\vspace@{\def\vspace##1{\crcr\noalign{\vskip##1\relax}}}
\def\multilimits@{\bgroup\vspace@\Let@
\baselineskip\fontdimen10 \scriptfont\tw@
\advance\baselineskip\fontdimen12 \scriptfont\tw@
\lineskip\thr@@\fontdimen8 \scriptfont\thr@@
\lineskiplimit\lineskip
\vbox\bgroup\ialign\bgroup\hfil$\m@th\scriptstyle{##}$\hfil\crcr}
\def\Sb{_\multilimits@}
\def\endSb{\crcr\egroup\egroup\egroup}
\def\Sp{^\multilimits@}

\newcommand{\be}[1]{\begin{equation}\label{#1}}
\newcommand{\ee}{\end{equation}}
\newcommand{\ba}[1]{\begin{eqnarray}\label{#1}}
\newcommand{\ea}{\end{eqnarray}}
\newcommand{\rf}[1]{(\ref{#1})}

\begin{document}
\author{U. G\"unther\dag\footnote{e-mail: u.guenther@htw-zittau.de},
A. Kuklov\ddag\footnote{e-mail: abk@bluebird.apsc.csi.cuny.edu}, and
A. Zhuk\S\footnote{e-mail: ai$\_$zhuk@imaff.cfmac.csic.es
\newline
on leave from: Department of Physics, University of Odessa, 2
Dvoryanskaya St., Odessa 65100, Ukraine}\\
\dag Gravitationsprojekt, Mathematische
Physik I,\\ Institut f\"ur Mathematik, Universit\"at Potsdam,\\ Am
Neuen Palais 10, PF 601553, D-14415 Potsdam, Germany\\
\ddag College of Staten Island, CUNY, NY, USA \\ \S Instituto de
Matem\'aticas y F\'{\i}sica Fundamental, CSIC, \\ C/ Serrano 121,
28006 Madrid, Spain
}

\title{Comment on stability of atoms and nuclei in
multidimensional TeV scale gravity}

\date{19.12.2000}

\maketitle

\begin{abstract}
We discuss stability of atoms and
nucleons in the presence of multidimensional gravity
characterized by the unified energy scale
~$M_{4+n}c^2\approx 1$ TeV. We point out that the
multidimensional gravitational attraction
can create additional bound states deep inside
of atomic and nucleon constituents. These states would be
characterized by sizes comparable to the TeV scale distance
~$R_T=\hbar/(M_{4+n}c)$. We show that
shallow gravity induced bound states between an electron and
a quark are ruled out, because
collapse processes in atoms to such bound states would occur
on time scales which are in contradiction to current data on the
stability
of ordinary matter.
The gravitational attraction may also produce
diquark states, which should be considered
in addition to those discussed in the literature.
The
estimates indicate that, for TeV scale gravity, the problem of
UV divergencies and UV renormalization is crucial.
Some universal
renormalization mechanism should exist, which
stabilizes atoms in the presence of
multidimensional gravity.
\end{abstract}\vskip0.5 cm

Recently the idea of multidimensional
TeV scale gravity has
been put forward in Refs. \cite{TEV} as a possible
resolution of the hierarchy
problem and as a step toward unification
of the Electro-Weak interactions and Gravity.
One of the main consequences
of this approach is
that the gravitational potential ~$\varphi_G(r)$~
produced by a point-like mass $M$ will deviate from the Newton's
law ~$\varphi_{NG}(r)= - GM/r$~
very significantly as
\begin{eqnarray}
\varphi_G(r)= - G{M\over r}\left({R_n\over r}\right)^n
\label{1}
\end{eqnarray}
\noindent at distances ~$r\ll R_n$. Here ~$R_n$~ stands for the
typical size of the $n$ internal dimensions of the
$(4+n)-$\-di\-mensional
spacetime and
$G$ denotes the observable
Newton's
gravitational constant.
The value $R_n$ is fixed by the $4-$dimensional Planck mass
~$M_{Pl}\equiv M_{(4)Pl}=\sqrt{\hbar c/G}$~ and energy
~$M_{Pl}c^2\approx
10^{19}$~GeV, and by the energy scale ~$M_{4+n}c^2$ of the
$(4+n)-$dimensional gravity. Specifically, for a toroidal internal space

\be{2}
R_n={\hbar \over c M_{4+n}}
\left({M_{Pl}^2 \over
M_{4+n}^2}\right)^{1/n}
=(2.0 \times 10^{-17}{\rm cm})
\left(\frac{1\ {\rm TeV}}{M_{4+n}c^2}\right)
\left({M_{Pl}^2 \over
M_{4+n}^2}\right)^{1/n}.
\ee
\noindent
Eq. (\ref{2}) indicates that for ~$M_{4+n}
\ll M_{Pl}$, the size of the compactified dimensions ~$R_n$
becomes much larger than the Planck  scale ~$L_{Pl}\equiv{\hbar / c
M_{Pl}}\approx 10^{-33}$cm. For example, in the case
$M_{4+n}c^2=1$ TeV \cite{TEV}, one finds ~$R_2\approx 10^{-1}$cm,
$R_3\approx 10^{-6}$cm, corresponding to
distances below the reach of present experimental tests of
gravitational forces.

In this letter, we point out that so large values of ~$R_n$~
imply that physics of atoms and nucleons (bound to a $(3+1)-$dimensional
world brane) becomes very sensitive to
the UV behavior of gravity in $4+n$ dimensions. Specifically, atoms and
nucleons may become unstable with respect to a collapse. This
can be avoided by introducing a cut-off for the potential (\ref{1})
below some distance $r_s$, determined by a specific screening or
renormalization behavior. Below we will
restrict the analysis to order of magnitude estimates
of the required values of $r_s$.

We start our consideration
by noting that a non-relativistic quantum mechanical
system with singular attractive
potential ~$\sim 1/r^{1+n}$~ and ~$n>1$ is known to have no
ground state \cite{LANIII}. The system is unstable and a collapse will
occur.
The collapse in a relativistic system
described by the Dirac equation with
potential ~$U(r\to 0)\sim 1/r^{1+n}$~ occurs as well
\cite{LANLIV}. Clearly, at atomic and
nuclear distances the strength of the additional gravitational
potential \rf{1}, e.g. between nuclei and shell electrons, is some
ten orders weaker than the strength of the nuclear Coulomb
potential\footnote{An estimate for Hydrogen ($Mc^2=0.94$ GeV,
$mc^2=0.51$ MeV ) and the fundamental scale $M_{4+n}c^2=1$ TeV
gives, e.g., at the Bohr radius $r_B$:
$\frac{\varphi_G(r_B)m}{\varphi_C(r_B)e}
=\frac{GmM}{e^2}\left(\frac{R_n}{r_B}\right)^n \approx
10^{-7}\times 4^n 10^{-9n}$ and at the proton surface ($r_p\approx
10^{-13}$cm): \ $\frac{\varphi_G(r_p)m}{\varphi_C(r_p)e} \approx
10^{-7}\times 10^{-4n}$.} $\, \varphi_C = Z |e|/r \, ,\, e < 0$ .
Even weaker it is compared with the strength of strong
interactions between constituent quarks inside the nucleons.
Nevertheless, regardless of the strength of the $1/r^{1+n}$ potential
(\ref{1}) at atomic or nuclear distances, a  collapse
will occur
if the form of the potential
(\ref{1})  holds  down to ~$r=0$ without lower bound.

A possible way to
exclude a collapse could consist in some
universal renormalization mechanism
which makes the potential
(\ref{1}) non-singular for ~$r\leq
r_s$. For our subsequent considerations we will
assume that such a renormalization is present.
We consider, for simplicity, the following effective
potential

\begin{eqnarray}
\tilde{\varphi}_G(r)= - G{M\over r+r_s}\left({R_n\over r +
r_s}\right)^n, \quad r_s>0,
\label{100}
\end{eqnarray}
\noindent
instead of (\ref{1}).

We assume that the cut-off follows from string theory/M-theory.
A natural scale for the upper energy cut-off is, then, the string
tension ~$M_sc^2$~. Thus, the distance ~$\hbar/M_sc$~ determines
the cut-off scale ~$r_s$. Obviously, such a cut-off leads to a lower
bound for
the discrete part of the energy spectrum.
In this regard, several situations for the gravity induced bound states
are possible: (i) there are "shallow" bound states characterized by
energy values $E=mc^2 - E_b$, such that $-mc^2 < E < mc^2$,
($E_b>0$ denotes the binding energy);
(ii) there are deep bound states only with $E<-mc^2$ residing in the
positron continuum; (iii)
no bound states exist in the potential (\ref{100}).

An instability of atoms due to large extra dimensions will occur
in the case (i), if the additional gravity induced bound states are empty.
Then a transition/collapse of an electron, e.g.,
from the K-shell to such a bound state (formed
inside the nucleus) can occur\footnote{Due to its
small "orbit" size, such a gravity induced bound state should be formed
between an electron and a nucleon constituent, i.e., a quark.
Clearly, the form
of the effective gravitating potential will deviate from \rf{100}
due to the mass distribution inside the nucleus. At atomic
distances of order of the Bohr radius the gravitating mass of the
center is given by the mass of the whole nucleus. Inside one of
the nucleons it is given by the mass of one of the constituent quarks
(e.g., for a u-quark, by $m_u\approx 310$ MeV \cite{con}),
whereas at smaller distances it
will be defined by the bare mass of the corresponding quark
($m_u\approx 4$
MeV). This means that a screening form factor should be included
into the potential \rf{100}. For our rough estimates we
do not take into account such a form factor.}.
In the case (ii), the bound
states will be filled by  electrons quite
rapidly from the vacuum, so that their partners --- positrons
are emitted to the outside (this process is similar to the generation of
a charged vacuum in the vicinity of overcritical nuclei \cite{ZP,GMR}).
As we will see below, these bound states are formed at distances
where the strength of the gravitational potential overcompensates
the Coulomb potential. Accordingly,
positrons will equally be attracted to practically
the same  bound states.
The filling of these
positron
states from the vacuum will be accompanied by the emission
to infinity of
their partners --- electrons. As a
result,
the system will contain real electron-positron pairs,
which are located in the vicinity of the gravitating center,
so that
the total charge of the nucleus does not change. In accordance with
the Pauli's exclusion principle, no
transitions to the filled deep bound states can occur from atomic
orbitals, so that the stability
of atoms is insured in this case.
Thus, the stability of atoms
is not affected at all by the presence of  deep
bound states. Obviously, no gravity induced transition/collapse
occurs in the case (iii).

In this letter, we consider possible
implications of the presence of  shallow
bound states with ~$-mc^2 <E < mc^2$, because,
as we have noted above, only in this case
the collapse problem becomes relevant.
In what follows we are confronted with the problem
of an adequate description of our system.
To be specific,
let us consider an atom.
If the collapse
occurs, e.g., for an electron from the K-shell
to one of the hypothetical gravity induced bound states
deep into the nucleus,
then the initial state is a well defined atomic state.
Due to its smallness
at atomic distances,
the additional gravitational potential can be taken into account
as perturbation, and
the corresponding non-relativistic low-energy physics is well
described by the Schr\"odinger equation.
The problem arises when one considers
the final state --- one of the hypothetical gravity induced bound
states.
We will see below, that if such states with energies $-mc^2 < E< mc^2$
indeed exist, then the corresponding potential energy and kinetic energy
of the bound electron would be of order of 1 TeV
or even higher.
Clearly, at such high energies the bound state approach
is not completely adequate\footnote{More adequate
could be a Bethe-Salpeter approach with inclusion of
electroweak interactions, and of the Higgs sector
as well as the of whole tower of massive Kaluza-Klein gravitons.
Additionally, at distances $r<R_T$ and energies above $M_{4+n}c^2$,
strong TeV scale quantum gravity effects, which could lead to topology
fluctuations, should also be taken into account.}.
We understand that the following estimates can only be interpreted as
a first crude indication of the collapse problem in TeV scale gravity.

In traditional models of heterotic string theory, the energy
~$M_sc^2$~ is only  few orders less than the Planck energy.
Accordingly, the cut-off distance $r_s$ is only slightly larger than
$L_{Pl}$
and much smaller than the characteristic TeV scale distance $R_T$:\ \
$L_{Pl}\sim 10^{-33}{\rm cm}\lesssim r_s \ll R_T\approx
10^{-17}$cm. In this case, as will be shown below,
the potential (\ref{100}) can support
plenty of the deep bound states with ~$E \ll -mc^2$.
Shallow bound states are possible as well.
Thus, the cases (i) and (ii) become possible.
We should also note that, if the number of
deep bound states is very large, the
gravitational potential may become so strongly modified
by the correspondingly large number of bound electrons and positrons, that
an absolute instability with respect to gaining
more and more particles from the vacuum develops.
The realization of the latter scenario
depends significantly on the ratio ~$r_s/R_T$.
In this letter, we will not discuss this issue in
detail. We only note that, in order to avoid the
vacuum instability, the
scale ~$r_s$~ must be much
larger than ~$10^{-33}$cm.
Thus, the
string energy must be lowered considerably.
Such a lowering of the string energy scale ~$M_sc^2$~ down to
1 TeV has been considered, e.g., in \cite{TEV,7226}. As discussed in
\cite{7226}, only certain types of strings can satisfy this
requirement without producing contradictions with observations.

A natural criterion for the existence of  bound
states in any attractive potential with a potential energy ~$U(r)$~
can be obtained by comparing it with the kinetic energy $T_{kin}$ of the
particle confined in a space region characterized by
some linear dimension ~$r$.
In general, no relativistic bound state exists if the condition
\begin{eqnarray}
|U(r)|\ll {\hbar c \over r}
\label{20}
\end{eqnarray}
\noindent
holds.
In TeV scale gravity, this condition (\ref{20}) is already
violated by the unrenormalized potential (\ref{1}) (produced by
the mass ~$M$~ and attracting the mass ~$m$~) at the distance
~$r_c$~ given by ~$|U(r_c)|\approx \hbar c/r_c$.
Employing ~$U(r)=\varphi_G(r)m$~
and excluding $R_n$ from (\ref{1}) by means of Eq. (\ref{2})
we find
\begin{eqnarray}
r_c\approx R_n \left(\frac{GmM}{\hbar c}\right)^{1/n}
=R_T \left(\frac{Mm}{M^2_{4+n}}\right)^{1/n}\, .
\label{5}
\end{eqnarray}
\noindent
The quantity ~$r_c$~ represents a rough estimate for the
size of the largest "orbit" of a gravity induced bound state in
the potential (\ref{1}). For ~$M_{4+n}c^2\sim 1$ TeV and $Mc^2=0.94$
GeV, ~$mc^2=0.51$ MeV, this gives the estimate $r_c\approx
10^{-10/n}R_T\approx 10^{-17-10/n}$cm.
{}From eq. \rf{5} we see that for particles with masses
~$Mc^2, mc^2 \le M_{4+n}$ the distance ~$r_c$ is smaller than the TeV
scale length $r_c \leq R_T$. This means
that the characteristic
size of the largest "orbit" lies already in the region of
strong TeV scale gravity where  (multidimensional) quantum
gravity processes become relevant.
These estimates confirm the relativistic ansatz for the kinetic energy
$T_{kin}\approx \hbar c/r$ of the bound particles which becomes much
larger than their rest energies. Accordingly, the rest masses $M$ and
$m$
in the potential
$U(r)=m\phi_G(r)$ should also be replaced by the corresponding
expressions for the
mass-energies ~$\hbar/(cr)$~ \cite{MTW}, so that effectively
~$U(r)=-(G/r)(\hbar/(cr))^2(R_n/r)^n$~ should be employed for
estimating
~$r_c$. As a result, one arrives at the relation $r_c\approx R_T$ \
\footnote{In this case, the condition ~$|U(r_c)|\approx \hbar c/r_c$~
can be written as ~$-\phi_G(r_c)/c^2\approx 1$~ which implies that the
gravitational potential is in the range of intermediate gravity and at
the
upper bound of the employed Newtonian approximation.}
for any  value of $n$.

The above analysis implies that the model is
very sensitive to the UV behavior induced by the extra dimensions.
Obviously, for
~$r_s \gg r_c$~ no bound states will be formed in the renormalized
potential (\ref{100}). In the opposite case, when  ~$r_s\ll r_c$,
the potential (\ref{100}) should contain plenty of
bound states. The
situation ~$r_s\sim r_c$~ is marginal.
It is, however, obvious that there is a certain critical
value ~$r'_s=\gamma r_c $, with ~$\gamma$~ being
some numerical factor of order of 1,
so that for ~$r_s> r'_s$~ the
bound states disappear, and for ~$r_s\to r'_s$~ from
below the bound states are shallow
and approach the
lower boundary of the upper part of the continuous spectrum.
In order to find ~$\gamma$, the exact problem
should be solved.

We note that the existence of empty shallow
bound states would strongly destabilize  ordinary
atoms on a time scale which is comparable with
times of usual atomic transitions. Indeed, let us
assume that there is a TeV scale gravity induced shallow bound
state ~$\psi_G({\bf r})$~ for an electron. In rough approximation,
such a state can be modeled by the Dirac equation with an effective
potential corresponding to a spherical well of a
depth ~$U_0\approx m|\tilde{\varphi}_G(0)|$~
and of the radius ~$r_s$. This
potential should transform into the standard
Coulomb potential
at ~$r> r_0\sim r_c$  \footnote{The distance $r_0$,
where
the potential energy of an electron
in the gravitational field of another particle
becomes comparable
with its energy due to the
Coulomb attraction by a unit electric charge, is in the
ultra-relativistic
limit roughly given by $(G/r_0)(\hbar/(cr_0))^2(R_n/r_0)^n\sim e^2/r_0$
(without running couplings taken into account). Accordingly, one has
$r_0\approx (\hbar c/e^2)^{1/(n+2)} R_T=(137)^{1/(n+2)} R_T>r_c\approx
R_T$.}.
It is important
that, regardless of the details of the problem
on distances ~$r\sim r_s$, the behavior
of   ~$\psi_G({\bf r})$~ outside the gravitational potential
is determined by the Coulomb potential and by
the value of the energy ~$E$. For shallow bound states
the long distance asymptotic is given by
\begin{eqnarray}
\psi_G({\bf r})=A_0{{\rm e}^{-\lambda r}\over r}\, ,
\label{A1}
\end{eqnarray}
\noindent
where ~$\lambda =\sqrt{m^2c^2-E^2/c^2}/\hbar$~ \cite{LANLIV};
~$A_0$~ stands for some constant
which should be determined from the
solution at small distances. To be more specific,
the solution in the region where gravity dominates
(~$r\sim R_T$~) must be matched with
the standard behavior in the Coulomb potential
\cite{LANLIV}. In some sense, the constant ~$A_0$~
defines the part of the gravity induced bound state
which spreads outside the region of strong gravitational attraction.
The quantity ~$\lambda^{-1} $~ sets a typical distance
scale for the extension of ~$\psi_G({\bf r})$.
Obviously, ~$\lambda^{-1} \gg r_c\approx R_T$~
for an electron.
The amplitude ~$A_0$~ in the asymptote (\ref{A1})
determines also the scattering phases
\cite{LANLIV}, and for any realistic potential
these phases should be finite (see \cite{LANIII}),
and therefore ~$A_0$~ should also be finite in the limit ~$r_s \to 0$.

The above analysis implies
that the overlap of the gravity induced shallow bound state
with any atomic state (with characteristic extension
~$r_{in}\approx 10^{-8}$cm) is determined by
the ratio ~$\chi =(r_{in}\lambda)^{-3} \geq
10^{-8}$ rather than by
the smallness ~$(r_c/r_{in})^3\sim 10^{-27}$.
For the K-shells of heavy
elements, the ratio ~$\chi$~ can be as large as $10^{-2}$.
Thus, the corresponding matrix elements of the
photon assisted
transitions from  atomic orbitals to
gravity induced shallow bound states are only
insignificantly reduced by ~$\chi$.
This becomes especially clear when
one considers observational implications of such transitions/collapses.
Indeed, in order to reconcile the existence
of the shallow bound states with
current data on the stability of
ordinary matter, the collapse rate
must be smaller than at least  the inverse
age of the Universe ~$10^{-18}$s$^{-1}$.
In fact,  observational limitations
on the stability are much stronger.
Experimental lower bounds on a
possible disappearance of K-shell
electrons in Iodine atoms yield, e.g.,
a half-life time of such processes
$ >10^{29}-10^{31}$s \cite{KSHELL}.
The
above value of $\chi$ can reduce the typical
rates of the atomic transitions from $10^{15}-
10^{8}$s$^{-1}$ to about ~$1-10^7$s$^{-1}$
only. This, implies that the existence of
shallow gravity induced bound states
(the possibility (i)) is
absolutely excluded for electrons.

Thus, if multidimensional TeV scale gravity
induces  bound states, then these states should
not be shallow. In fact, these states should lie
deep enough so that the escaping part of the
electron wave function (\ref{A1})
does not significantly modify
the probability of finding the
atomic electrons at the nucleus.
Otherwise, such deep
gravity induced  states would
contradict to the current data on the
atomic hyperfine
splitting, which is known to be sensitive to
the probability of finding electron near the
nucleus \cite{LANIII}.
In this  letter, we will not further discuss possible
implications of the hypothetical gravity induced deep bound states.
We only note that in the case (ii),
the Pauli exclusion principle would
stabilize the system, so that transitions of atomic electrons
to such states will not occur.
Thus, in the case (ii), the presence of the
TeV scale gravity potential will have no direct consequence
on the stability of atoms.

We note that, for a realization of the case  (iii) with no bound
states present at all, the cut-off ~$r_s$~ should satisfy the condition
\begin{eqnarray}
\displaystyle r_s>\gamma r_c \approx R_T .
\label{81}
\end{eqnarray}
\noindent

The above analysis has been carried out for electrons in the field
of some massive center. A similar stability problem, which, however,
leads to less drastic consequences than in the case of the
atomic electrons,
exists for
quarks in  nucleons. In this case,
the TeV scale gravity may induce, e.g.,  diquark states at various energy
scales. Such states, which are characterized by
the quarks separation ~$\sim R_T$~  should be considered in addition
to those discussed  in the literature
\cite{dq1}.
We will not dwell on this here.

Summarizing the above discussion, we are led to the conclusion that
the UV problem is very acute in
multidimensional TeV scale gravity\footnote{Similar
stability arguments for atoms
can be applied to the physics on a brane
of some finite width $R_B$. The
traditional attractive Coulomb potential
should convert into its
multidimensional counterpart
~$V_C(r)\sim 1/ r^{n+1}$~ at ~$r\lesssim R_B$.
Accordingly, additional induced states with ~$r_c \sim R_B$~
could exist unless some universal renormalization
mechanism in ~$4+n$~ dimensions makes the potential
less singular at such distances.},
and any renormalization scheme
should ensure that at least no shallow
bound states with  binding energies below
the electron rest mass can be formed.
A similar stability problem as for TeV scale gravity will also
occur in Randall-Sundrum models \cite{RS}.
Although the effective on-brane gravitational potential on
the Planck brane differs from that on a probe brane, the short
distance behavior on both types of branes
is given by a $r^{1+n}$ term with $n\ge 1$ \cite{RS2},
so that an appropriate
circumvention of the corresponding collapse  problem should be found.
We should also note that our results are not valid
for models \cite{DG}, where due to the induced metric on the brane,
the gravity on the brane is exactly $4-$dimensional all the way down
to distances of order ~$\hbar/M_{Pl}c$.

\bigskip
{\bf Acknowledgments} We thank Valery Rubakov and Gia Dvali for
valuable comments. This work
has been initiated during A.Z.'s visit at the College of Staten
Island of CUNY. A.Z. acknowledges support by CUNY, the US
Federal Department of Foreign Affairs,
the programme SCOPES
(Scientific co-operation between Eastern Europe and Switzerland) of
the Swiss National Science Foundation, project No. 7SUPJ062239,
and the Spanish Ministry of Education, Culture and Sport. U.G.
acknowledges financial support from DFG grant
KON 1575/1999/GU522/1.

\end{document}